\documentclass[12pt]{article}
\usepackage{amssymb,amsfonts,epsfig}

\renewcommand{\theequation}{\arabic{section}.\arabic{equation}}
\newcommand{\gapprox}{%
\mathrel{%
\setbox0=\hbox{$>$}\raise0.6ex\copy0\kern-\wd0\lower0.65ex\hbox{$\sim$}}}
\newcommand{\be}{\begin{equation}}
\newcommand{\ee}{\end{equation}}
\newcommand{\bea}{\begin{eqnarray}}
\newcommand{\eea}{\end{eqnarray}}
\newcommand{\bref}[1]{(\ref{#1})}

\def\ie{{\it i.e. }}

\def\t{\theta}

\def\del3{\delta^{(3)}}

\def\undos{{1 \over 2}}

\def\otaula{\begin{tabular}}
\def\ctaula{\end{tabular}}
\def\espai{\;\;\;\;\;\;}

\def\a{\alpha}
\def\b{\beta}

\def\hg{\hat{g}}

\def\l{\lambda}

\def\s{\sigma}

\def\avall{\vspace{1cm}}

\def\ka{K\"ahler }

\def\T{\Theta}
\def\CN{$\mathcal{N}$}

\def\nun{N\'u\~nez }
\def\CF{\mathcal{F}}
\def\cC{\mathcal{C}}
\def\tPhi{\hat{\Phi}}
\def\MR{\mathbb{R}}

\def\cO{\mathcal{O}}
\def\hA{\hat{A}}
\def\hF{\hat{F}}
\def\hg{\hat{g}}

\begin{document}

\begin{titlepage}

\setcounter{page}{0}
\begin{flushright}
ECM-UB-02/17 \\
\end{flushright}

\vspace{5mm}
\begin{center}
{\Large {\bf Supergravity Dual of Noncommutative \CN=1 SYM }}
\vspace{10mm}

{\large Toni Mateos$^a$, 
Josep M. Pons$^a$ and Pere Talavera$^{a,b}$} \\
\vspace{5mm}
$^a${\em 
Departament d'Estructura i Constituents de la Mat\`eria,
Universitat de Barcelona,\\
Diagonal 647, E-08028 Barcelona, Spain.}\\[.1cm] 
$^b${\em 
Departament de F{\'\i}sica i Enginyeria Nuclear,
Universitat Polit\`ecnica de Catalunya,\\
Jordi Girona 31, E-08034 Barcelona, Spain.}
\vspace{5mm}

\vspace{20mm}

\centerline{{\bf{Abstract}}}

\end{center}
\vspace{5mm}

We construct the noncommutative deformation of the Maldacena-\nun
supergravity solution. The background describes a bound state
of  D5-D3 branes wrapping an $S^2$ inside a Calabi-Yau three-fold, and
in the presence of a magnetic $B$-field. The dual field theory in the IR
is an \CN=1 $U(N)$ SYM theory with spatial noncommutativity.
We show that, under certain conditions, the massive Kaluza-Klein states 
can be decoupled and  
that UV/IR mixing seems to be visible in our solution.
By calculating the quark-antiquark potential via the Wilson loop 
we show confinement in the IR
and strong repulsion at closer distances. We also compute
the $\b$-function and show that it coincides with the 
recently calculated commutative one.

\vfill{
 \hrule width 5.cm
\vskip 2.mm
{\small
\noindent E-mail: tonim,pons,pere@ecm.ub.es 
}}

\end{titlepage}

\newpage

\section{Introduction and Conclusions}
\setcounter{equation}{0}

Since the appearance of the AdS/CFT correspondence
\cite{Maldacena:1997re,Gubser:1998bc,Witten:1998qj}
many efforts have been devoted to obtain supergravity duals of gauge
theories with less than maximal supersymmetry.
One possible way to achieve this is by wrapping D-branes in supersymmetric
cycles, and many examples of these have been constructed,
see $e.g.$ \cite{Maldacena:2000mw}-\cite{Naka:2002jz}.
Many 
qualitative as well as some quantitative aspects of the perturbative and non-perturbative
gauge theories have been extracted from the supergravity duals.

Similarly, the observation that noncommutative (NC) field theories arise as a low-energy
limit of string theory \cite{sei-wit} in a $B$-field background
motivated the construction of gravity duals for such theories 
\cite{malda-russo,hashimoto-itzhaki}. Many duals of maximally
supersymmetric NC theories in diverse dimensions are known (see
\cite{larsson} for a summary).

Trying to join both ideas, the authors of \cite{ub-grup} constructed
the dual of an \CN=2 NC $U(N)$ gauge theory in three dimensions by
wrapping D6 branes in a supersymmetric four-cycle, and turning on
a $B$-field in the flat directions of the brane. The method relies on
proposing an ansatz in 11d supergravity, obtaining the Killing spinors,
and solving the first order BPS equations. They also showed that,
unlike most cases considered in the literature,
the solution could not have been obtained in the
corresponding gauged supergravity.

One of the most relevant supergravity duals was constructed
by Maldacena and \nun (MN) in \cite{mn}. Their background is dual
in the IR to an  \CN=1 $U(N)$ pure SYM theory in four dimensions.
Among other properties, the field theory is confining
and exhibits chiral symmetry breaking. Recently \cite{lerda}, the complete
perturbative NSVZ $\b$-function has been extracted from the supergravity solution,
giving further evidence for the duality.

In this paper we construct the supergravity dual of a noncommutative 
\CN=1 pure SYM theory in four dimensions by deforming the MN solution
\cite{mn} to incorporate a magnetic $B$-field along two spatial flat directions
of the branes. Our background represents a bound state of D5-D3 branes,
with the D3 smeared in the world-volume of the D5. Both types of
branes are wrapping an $S^2$, so that their world-volume
can be summarised in an array as follows
\begin{center}
\[
\begin{array}{c | c c c c c c}
{\rm IIB} &x^0 & x^1 & x^2 & x^3 & \t & \phi  \\ \hline
{\rm D5}& -&-&-&-&-&- \\
{\rm D3}&-&- & &&-&-\\
B_{[2]} & & &- &-&&\\
\end{array}\]
\end{center}
We show that, unlike the commutative case, the parameters of our solution can
be chosen such that the curvature and the dilaton remain small everywhere, without
need to perform an S-duality. Furthermore, the massive KK modes of the $S^2$
can be decoupled by appropriately choosing our parameters.
We also check that in the deep IR limit, the background does not completely
reduce to its commutative counterpart, in part due to the non-vanishing of
the RR field of the D3 branes. This is in contrast to the dual
of \CN=4 NCSYM and it is
presumably due to the fact that UV/IR mixing is absent for \CN=4, but
it is expected to be present in cases with less supersymmetry.

We proceed to extract gauge theory physics from the supergravity solution
and study the quark-antiquark potential, by using the proposal of \cite{malda-wilson}
relating closed Wilson loops in the field theory with minimal string worldsheets
in supergravity. We find that at large distances, the potential is linear, 
with exactly the same slope as in the commutative case. Nevertheless,
as we bring the quarks closer, the potential becomes strongly repulsive. This
is in agreement with the expected results. In the IR, our 
metric and
$B$-field agree with the commutative result, and these are the only objects
to enter in the calculation of the Wilson Loop, so although our solution
differs in other respects from the commutative one, the slopes should
coincide.
But as the quarks approach,
the noncommutative uncertainty relation seems to dominate and the force
becomes repulsive.

Finally we adapt the methods of \cite{lerda, zaffaroni} to the noncommutative
background, and calculate the $\b$-function. The ordinary AdS/CFT identification
between field theory operators and supergravity fields gets more complicated
in NC theories, due to the fact that they do not admit any local
gauge invariant operators. Nonetheless, they do admit invariant operators
local in momentum space which involve open Wilson lines
\cite{gross,rey}. Recent computations seem to indicate that 
the supergravity fields act as sources of such kind of operators
\cite{liu1,okawa,das}. In any case, the fact that our supergravity
fields do not depend on the noncommutative coordinates makes it easier
to obtain the gravity fields that are dual to the gauge theory operators.
This is because they do not carry momentum along the NC directions and
therefore they must couple to zero-momentum gauge-invariant operators of the field
theory.
As we further develop in section 4, we can identify the supergravity field dual
to the gaugino condensate, along the same lines as in \cite{lerda, zaffaroni}.
The result is that the NCYM $\b$-function matches completely 
the one extracted from the commutative background.
This is plausible from a field theory point of view, where
it is known that, at least perturbatively, the $\b$-function of a NC $U(N)$
gauge theory
is the same as the one of a commutative $SU(N)$ one\footnote{Note that
the write $U(N)$ instead of $SU(N)$ because the NC $U(1)$ degrees of freedom  
do not decouple in NC-theories \cite{Armoni}.}(see 
\cite{Armoni,Martin,Hayakawa}, as well as the review \cite{szabo} and references therein).

The paper is organised as follows: in section 2 we review in a generic manner 
the construction of the deformed noncommutative theories, and we apply it to the case 
presented in  \cite{mn}. We include an analysis of the supergravity validity and
the decoupling of the Kaluza-Klein modes, comparing
with the commutative case. We then turn to analyse the IR/UV mixing, checking explicitly
that in the deep IR the noncommutative background does not reduce completely to the commutative one.
In section 3 we compute the quark-antiquark potential via Wilson loops and discuss
the peculiar fine tuning of the string velocity that is required.
In section 4 we evaluate the field theory $\beta$-function,
showing its agreement with the commutative one. In order to make this letter more readable,
we leave for the appendix some conventions and definitions that we use.

We shall work in units where $\alpha'=1$ and restore it by dimensional
analysis only when needed.

\section{Construction of the Supergravity Duals}
\setcounter{equation}{0}

In this section we construct the supergravity dual of a noncommutative
$U(N)$ field theory with \CN=1 in four dimensions. We 
shall start from the remarkable commutative solution that 
Maldacena-\nun 
obtained by 
making use of the previously found nonabelian monopole solution of \cite{volkov1,volkov2}, 
and deform it
to incorporate a magnetic $B$-field along the directions of the brane.

\subsection{The method}

There are several different ways to obtain such deformations. 
In \cite{ub-grup} , the dual of a noncommutative theory with \CN=2 in three
dimensions was constructed by making an ansatz at the level of D=11 supergravity and
solving the BPS equations. One of the advantages of such  procedure
is that it allows one to obtain the Killing spinors of the background.
They can be used to confirm that one truly has a Dp-D(p-2) non-threshold 
bound state, that no supersymmetries are lost by turning on the $B$-field,
to build the generalised calibration, and to show that the solutions
could not have been found in the corresponding gauged supergravity.

Another way to construct the deformation is by noticing that diagonal
T-dualities typically produce Dp-D(p-2) bounds states with a background
$B$-field. This idea was proposed in \cite{myers, Costa} before the 
understanding of noncommutative field theories as low-energy limits of string
theory, and the technique has been greatly improved in the past two years.
Let us briefly review how the original and the improved methods work, and
why they are equivalent.

Suppose we have a Dp-brane in flat space along the 
directions $\{x^0,x^1,...,x^p\}$. We would like to perform
a T-duality along a diagonal axis in the plane $(x^p,x^{p+1})$. Equivalently,
we can rotate the brane in that plane and simply T-dualise along $x^{p+1}$.
In the last picture, the originally tilted brane had coordinates satisfying
\be
\partial_{n}\left(x^p + \tan\t \,  x^{p+1}\right)=0\,, \espai\espai
\partial_{t}\left(x^p - \cot \t \, x^{p+1}\right)=0
\ee
where $\partial_n$ and $\partial_t$ are normal and tangent derivatives with
respect to the string worldsheet's boundary, and $\t$ is the angle of rotation. Now, 
T-duality along $x^{p+1}$ exchanges Neumann and Dirichlet conditions and produces
\be
\partial_{n} x^p + \tan\t \, \partial_{t} x^{p+1}=0\,, \espai\espai
\partial_{n} x^{p+1} - \tan \t \, \partial_t x^{p}=0\,.
\ee
But this mixed boundary conditions can be interpreted as those of a 
string attached to a $D(p+1)$ brane in the presence of a $B$-field 
\be
\partial_n x^{\mu}\, - \,\CF^{\mu}~_{\nu} \, \partial_t x^{\nu}\,=\,0
\ee
where $\CF_{[2]}=B_{[2]}+2\pi \a' F_{[2]}$ and, in this case, we
have induced $\CF_{12}=-\tan\t$. Such gauge invariant field strength
produces D(p-1) charge in the world-volume of the D(p+1) through
the Wess-Zumino term.

This is, {\it grosso modo}, the original method proposed in \cite{myers, Costa}, 
where it was applied to several cases of branes in flat space. Nevertheless,
it still had some technicalities that made it difficult to generalise to
branes in other backgrounds. Maybe the most relevant was that T-dualities
need to be performed along isometries. In our case, the T-duality was
performed along a diagonal direction involving one coordinate along the
brane and one transverse to it, and this is not
an isometry of the supergravity solution. This was originally solved by 
delocalising the Dp branes along
the $x^{p+1}$ axis, for example by adding an infinite number of parallel branes.
In the supergravity solution of flat p-branes, this just amounts to 
changing slightly the form of the harmonic function $H(r)$: instead of being
harmonic in the whole transverse space of dimension $10-p-1$, one can choose it
to be harmonic in one dimension less, \ie in a $10-p-2$ space. Schematically,
\be {\rm Dp \,\,localised:} \espai
H(r)=1+{1\over r^{7-p}}\, , \espai r^2=\sum_{i=p+1}^{10} (x^i)^2\,. \ee\be 
{\rm Dp \,\,delocalised \,\, in \,\, x^{p+1}:} \espai
H(\tilde{r})=1+{1\over \tilde{r}^{6-p}}\, , \espai \tilde{r}^2=\sum_{i=p+2}^{10} (x^i)^2\,.
\label{replace}
\ee
As can be seen, delocalising a brane is fairly simple when we are in flat
space and we know the whole geometry solution. The difficulty would increase
if we were only given the near horizon region. There, the  harmonic function
can be very hard to recognise. Indeed, if we also abandon flat space backgrounds,
the transverse space to the brane is typically a sophisticated fibre bundle, and
a better method to delocalise the brane is needed.

The way this can be achieved is just by starting with a brane of one dimension
higher, say a D(p+1) along $\{x^0,...,x^{p+1}\}$ and by T-dualising along $x^{p+1}$. 
Since the change from Neumann to Dirichlet boundary condition of $x^{p+1}$
does not fix its zero mode, the resulting Dp brane will not be localised in the
$x^{p+1}$ axis. In the supergravity dual, one just needs to use the T-duality
rules to transform the closed string background. In flat space, it is easy to 
check that this is equivalent to the replacement \bref{replace}, no matter if we
started with the whole geometry or just the near-horizon.

The last refinement of the original method consists on substituting the rotation
of the delocalised brane by a more mechanical algorithm. It just exploits the fact
that rotating the brane is equivalent to: first T-dualising one of the world-volume
directions, then turning on a constant $B$-field, and then T-dualising back.

Therefore, the improved method for producing the noncommutative configurations can 
be summarised, from a supergravity point of view, as follows:

\avall

{\bf (i)} Start with a supergravity solution of a Dp along $\{x^0,...,x^p\}$. We require
that at least two of these directions, say $\{x^1,x^2\}$,  are flat, while the others
may or may not be wrapped along any compact cycle. We compactify $x^1$ and $x^2$ on
a torus so that $\partial_{x^1}$ and $\partial_{x^2}$ generate circle isometries.

{\bf (ii)} T-dualise along $x^2$. This produces a D(p-1) brane delocalised along $x^2$.

{\bf (iii)} Rotate the D(p-1) along the $(x^1,x^2)$ plane by T-dualising along $x^1$,
turning on a constant $B$-field $B=\T\, dx^1 \wedge dx^2$, and T-dualising along $x^1$ again.
The introduction of $B$-field does not modify the equations of 
motion because its field strength is zero, and the Chern-Simons term of the
corresponding supergravity Lagrangian is a total derivative.

{\bf (iv)} T-dualise back on $x^2$. This is the diagonal T-duality 
of a delocalised and rotated brane that we mentioned. It produces 
a bound state of Dp-D(p-2) in the
background of a non-trivial $B$-field. Finally, uncompactify $\{x^1,x^2\}$
by sending the radii of the torus to infinity.

\avall

Supersymmetry is preserved throughout this procedure if the spinors originally
did not depend on $x^1$ and $x^2$ \cite{ortin}, as is typically the case. The introduction
of the $B$-field in step $(iii)$ does not break supersymmetry either, since only
$H_{[3]}=dB_{[2]}=0$ appears in the supersymmetry variations of supergravity.

We conclude this section by making a few remarks about the improved method. 
The first is that it generalises easily to include $B$-fields with rank 
higher than two. The second is the non-trivial fact that, as pointed out
in \cite{sundell}, when the $B$-fields are magnetic, the following flow diagram
holds

\begin{figure}[h]
\begin{center}
\includegraphics[width=12cm,height=10cm]{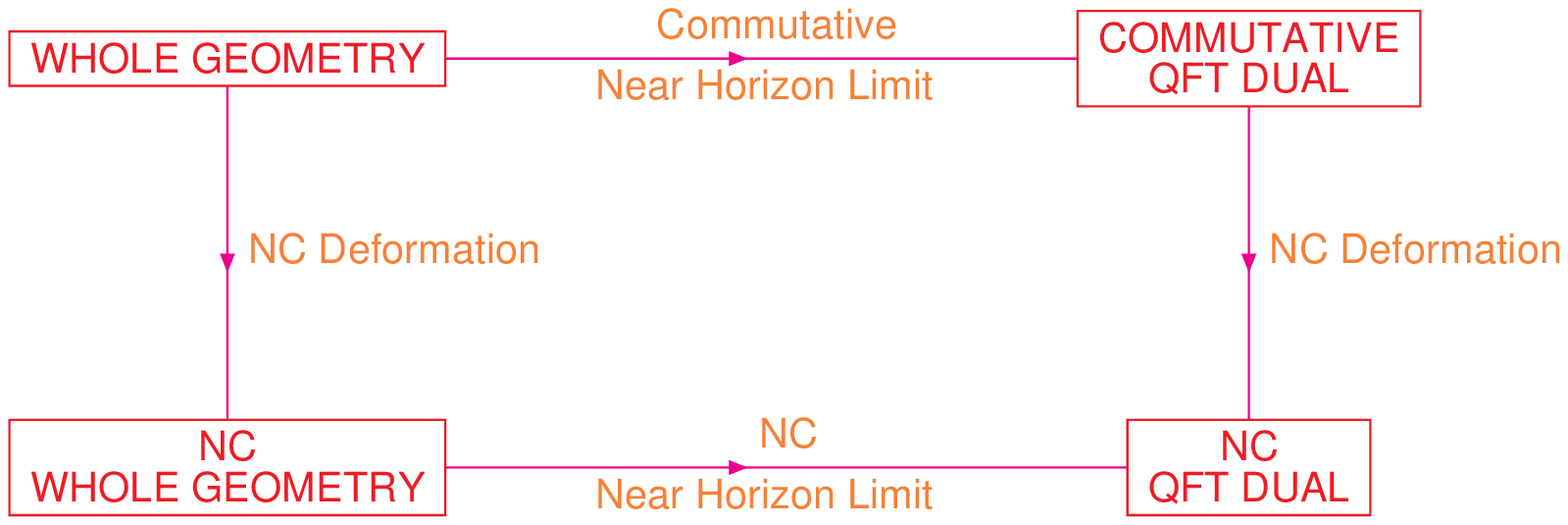}
\vspace{-4cm}
\end{center}
\end{figure}

This is crucial for our purposes, since in the  supergravity duals of 
wrapped branes one only knows the near-horizon region. The third and last remark is that
the improved method has been widely used to obtain duals of maximally supersymmetric
field theories, as in \cite{malda-russo,hashimoto-itzhaki,larsson,cederwall},
and of \CN=2 as in \cite{polchinski}.

\subsection{Supergravity dual of Maldacena-\nun }

Once the method has been discussed, we present the results of the noncommutative
deformations of one of the most relevant supergravity duals obtained in the
literature.
The solution \cite{mn} is dual in the IR to
\CN=1 pure SYM in four dimensions.
It represents a stack of $N$ D5 branes wrapping
an $S^2$ inside a Calabi-Yau three-fold. The solution reads
\be
\label{metrica}
ds^2_{IIB}=e^{\Phi} \left[  dx^2_{0,3}+
N \, \left( d \rho^2+ e^{2g(\rho)} d\Omega_2 + {1 \over 4} (w^a-A^a)^2 \right)
\right]\,,
\ee
\be F_{[3]}=dC_{[2]}={N\over 4}\left[-(w^1-A^1)\wedge(w^2-A^2)\wedge(w^3-A^3)
+ \sum_{a=1}^3 F^a\wedge(w^a-A^a)\right]\,,
\ee
\be \label{com-dilato}
e^{2\Phi}=e^{2\Phi_0}\,\,{\sinh 2\rho \over 2 e^{g(\rho)}}
\ee
where the definitions of the quantities appearing above are written
in the appendix. We skip the intermediate steps and give the result of
the noncommutative deformation of this background with a magnetic $B$-field along
the $\{x^2,x^3\}$ plane,
\be \label{nc-mn}
ds^2_{IIB}=e^{\Phi} \left[ dx^2_{0,1}+
 h^{-1} dx_{2,3}^2 +
N \, \left( d \rho^2+ e^{2g(\rho)} d\Omega_2 + {1 \over 4} (w^a-A^a)^2 \right)
\right]\,,
\ee
\be \label{nc-dilato}
F_{[3]}=dC_{[2]}= {\rm unchanged}\,,
\espai\espai
 e^{2\tPhi}= e^{2\Phi} h^{-1}\,,
\ee
\be
B_{[2]}=-\, \Theta \, {e^{2\Phi} \over h} \,dx^2 \wedge dx^3\,, \espai\espai
C_{[4]}=\Theta \, {e^{2\Phi}\over 2h} \, C_{[2]} \wedge dx^2 \wedge dx^3
\ee
where we defined
\be \label{last}
h(\rho)=1+\Theta^2 e^{2\Phi}\,.
\ee
Notice that we use $\tPhi$ for the new value of the dilaton and $\Phi$ for the
one appearing in \bref{com-dilato}. Also, $\T$ is the noncommutative 
parameter, while $C_{[2]}$ and $C_{[4]}$ are the type IIB 
Ramond-Ramond potentials, with field strengths $F_{[3]}$ and $F_{[5]}$
respectively. Note as well that the $B$-field is not trivial
(its field strength is non-zero) but it is constant along the directions
of the brane.

A few remarks concerning (\ref{nc-mn}-\ref{last}) are in order. First of all, notice 
that in the commutative limit $\T \rightarrow 0$ we have $h(\rho)\rightarrow 1$
and hence we smoothly recover the whole commutative background of MN.
Second, the solution describes a bound state of D5-D3 branes, 
with the D3 smeared in the world-volume of the D5, and partially
wrapped in the two-sphere. If we denote by 
$(\theta,\phi)$ the coordinates of the $S^2$ in \bref{nc-mn}, and
by $(\t_1,\phi_1,\psi)$ the ones of the transverse $S^3$, we can 
summarise the configuration in the following array
\begin{center}
\[
\begin{array}{c | c c c c c c c c c c}
{\rm IIB} &x^0 & x^1 & x^2 & x^3 & \t & \phi & \rho & \t_1 &\phi_1 & \psi \\ \hline
{\rm D5}& -&-&-&-&-&-&&&& \\
{\rm D3}&-&- & &&-&-&&&&\\
B_{[2]} & & &- &-&&&&&&\\
\end{array}\]
\end{center}
Third, as in the MN solution, the metric is completely
regular at the origin.


\subsection{Validity of Supergravity and KK states}
\label{sugra}

\begin{figure}[t]
\begin{center}
\includegraphics[width=6cm,height=4cm]{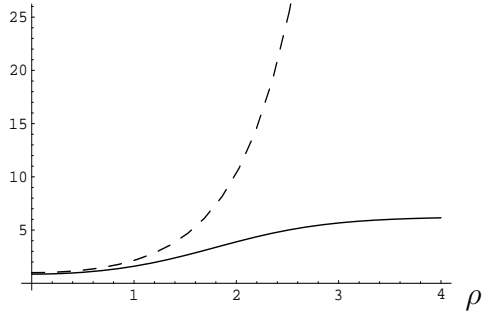}
$\rho$
\end{center}
\caption{Dilaton behaviour as a function of the transverse coordinate.
The full line corresponds to $e^{2\hat{\Phi}(\rho)}$ (NC case)
and the dashed line to $e^{2\Phi(\rho)}$ (commutative case). While the former
remains finite at any value of the variable $\rho$, the latter blows up at infinity.}
\label{figdila}
\end{figure}

Before continuing with our discussion,
let us analyse the conditions for the NC-MN solution to be a valid approximation
of string theory. The main difference with respect to the commutative solution
is that the dilaton does not diverge at the boundary, due to the factor $h^{-1}$ in
\bref{nc-dilato}. It acquires its maximum value at infinity -see fig. \bref{figdila}-, 
where $e^{\hat\Phi}\rightarrow \T^{-1}$.
So if we want to keep small everywhere the corrections coming from 
higher order diagrams of string theory, we just need to demand 
\be \label{condi}
\T\gg 1\,.
\ee

The second validity requirement comes from the curvature. In the noncommutative
geometry \bref{nc-mn}, the scalar of curvature ${\cal R}$ vanishes at infinity and it
acquires its maximum value at the origin. Requiring the curvature to be small everywhere
implies explicitly 
\be \label{curva}
\vert {\cal R}\vert_{\rm max}\, =\, \vert {\cal R}(\rho = 0)\vert \,
= \, \frac{32}{3N}\,\frac{e^{-\Phi_0}}{ \left(1+ \T^2e^{2\Phi_0}\right)}
\, \ll \, 1\,.
\ee
In order to obtain a truly pure \CN=1 NCSYM at low energies, 
conditions \bref{condi} and \bref{curva} should be compatible with the 
decoupling of the massive Kaluza-Klein modes of the wrapped $S^2$.
Since the only change in the metric with respect to the commutative 
one is in the $(x^2,x^3)$-plane, the KK modes decoupling condition
is exactly the same as in \cite{mn}, namely 
\be \label{kk-res}
N e^{\Phi_0}\ll 1 \,.
\ee
It is easy to show that the three inequalities \bref{condi}-\bref{kk-res}
can be satisfied simultaneously if we choose the three parameters $N, \Phi_0$ and
$\T$ to verify
\be \label{final-condi}
\frac{e^{-3\Phi_0}}{\T^2}\,\ll \, N \, \ll e^{-\Phi_0} \ll \T \,.
\ee
We shall see in section 3 that a further
restriction will have to be imposed in order to study the quark-antiquark potential.

\subsection{Properties of the solution and UV/IR mixing}

As we mentioned already, the NC-MN solution \bref{nc-mn} reduces to the commutative one
when we send $\T$ to zero. This corresponds to the fact that, classically,
noncommutative theories reduce to commutative ones in this limit.
This remark does not hold
quantum-mechanically and constitutes
one of the most interesting facts of the NC field theories, which is
 related to the so-called UV/IR mixing
\cite{minwalla,susskind}.
Perturbatively, the Feynman diagrams can be divided in planar and non-planar
ones. Planar diagrams give identical contributions (up to external phases) as
the corresponding commutative ones. The change comes from the non-planar diagrams,
which are regulated by the noncommutative phase and typically 
diverge when one sends the external momenta or the noncommutative parameter to 
zero. In other words, very high-energy modes seem to strongly affect the IR 
physics.

Exceptionally, this phenomenon is known to be absent in the superconformal \CN=4 SYM
\cite{szabo} due to the lack of UV divergences in the underlying
commutative theory. This was confirmed in the supergravity dual 
\cite{malda-russo,hashimoto-itzhaki} by observing that their configuration 
reduces in the deep IR to the usual $AdS_5 \times S^5$.

Let us devote our attention to review the metric and the field content
of the noncommutative case and carefully analyse if it reduces
to its commutative counterpart in the deep IR.
Thus we are interested
in the $\rho\rightarrow 0$ limit of \bref{nc-mn}. The key observation
is that the function $h(\rho)$ tends to the constant value $h(0)=1+\T^2e^{2\Phi_0}$.
Thus the coefficient multiplying the noncommutative coordinates
 $dx_2^2+dx_3^2$ becomes a \emph{constant}, which
could have been absorbed in a rescaling of the coordinates from the very beginning
\be
\hat{x}^i = {x^i \over h(0)}\,, \espai\espai\espai i=2,3\,.
\ee
We would like to clarify that although the NC metric seems to tend to the commutative
one, its derivatives do not. This can be easily inferred for example from the
value of the scalar of curvature at $\rho=0$ \bref{curva}, which does depend on $\T$.
In general, all objects constructed from derivatives of the metric may differ
from their commutative counterparts. The same observation applies to the following
analysis for the rest of the fields.

Consider now the $B$-field, which
tends to a constant in this limit, and so it becomes pure gauge. We could have started from
the beginning with a gauge-related $\hat{B}$-field
\be
\hat{B}_{[2]} = B_{[2]} + d\left(  {\T\over \T^2 +e^{-2\Phi_0}} \, x^2 \, dx^3\right)
\ee 
that would vanish in the deep IR.
In any case, the gauge-invariant field strength $H_{[3]}=dB_{[2]}$  vanishes 
for $\rho \rightarrow 0$. Let us now analyse the dilaton. In this limit, we
obtain
\be
e^{2\tPhi} \, \longrightarrow \,{e^{2\Phi_0} \over h(0)}
\ee
which just amounts to a redefinition of the value of the dilaton at the origin.
Furthermore, the field strength $F_{[3]}$ that couples magnetically to the D5
is unchanged everywhere. 

All fields considered so far reduce to their commutative counterparts
 (although, as we have said, the entire geometry does not).
Let us now analyse the remaining $C_{[4]}$ that couples to the D3 branes.
It is easy to see that in the deep IR limit it does not vanish. Since
this is not a gauge-invariant statement, we can look at its field strength\footnote{Indeed, one should make it 
self-dual by defining $\tilde{F}_{[5]}= \undos ( F_{[5]}+ *F_{[5]} )$ but
the following discussion is not affected. Signs are chosen according to
the conventions of \cite{myers}.},
\be
F_{[5]}=dC_{[4]}-\undos\left( B_{[2]}\wedge F_{[3]} -C_{[2]}\wedge H_{[3]}\right)\,.
\ee
Upon substitution we obtain $F_{[5]}=-B_{[2]}\wedge F_{[3]}$,
which tends to
\be
F_{[5]} \longrightarrow -{\T \over e^{-2\Phi_0} +\T^2}\,\, dx^2\wedge dx^3 \wedge F_{[3]}\,.
\ee
when $\rho \rightarrow 0$ and, therefore, does not vanish.
Indeed, this statement is still
coordinate dependent. One way to make it more rigorous is to construct
scalar quantities out of $F_{[5]}$. One could for example compute
$F^2_{[5]}$ where all indices are contracted with the inverse metric.
Performing this calculation in the \CN=4 duals of 
\cite{malda-russo,hashimoto-itzhaki}, where the configuration corresponds
to D3-D1 bound states instead of D5-D3, one finds that the D1 field
strength vanishes quickly in the IR, while the D3 one remains finite.
Furthermore, one could compute it as well in the case of D5-D3 in flat space,
or more generally in the rest of Dp-D(p-2),
directly from \cite{myers}. The result is again that the lowest brane
field strength vanishes at the origin, while the one of the Dp remains.
Nevertheless, in our case, the square of $F_{[5]}$ remains constant
too, so that the D3 field strength does not vanish!
Therefore,
in the deep IR limit, all fields reduce to the commutative result
except for $F_{[5]}$. This difference is originated in the fact that
the MN metric is completely regular at the origin.

Presumably,
this could be a signal of the UV/IR mixing that is expected to occur
in \CN=1 and \CN=2 theories. We should mention here
 the observation in \cite{susskind}
that in the large $N$, non-planar diagrams are sub-leading with respect
to the planar ones, so that noncommutative effects should not be
visible. Even if this was true, we recall that our solution does not
necessarily require to send $N$ to infinity, so that it is reasonable
to see a different IR behaviour from the commutative case.

\section{Quark-antiquark potential}
\setcounter{equation}{0}

In this section we obtain the quark-antiquark potential in the \CN=1 $SU(N)$
field theory by examining the behaviour of the Wilson loop. 
We proceed as usual by setting a stack of
$N$ D-branes at the origin and pulling out one of them to the boundary.
We then consider an open string whose boundary on the probe brane describes
a rectangular Wilson loop, with one time-like and one noncommutative direction.

\subsection{Evaluation of the Wilson loop}

In the case at hand the Wilson loop average is obtained \cite{malda-wilson} 
by minimising the Nambu-Goto action
in the presence of the $B_{[2]}$ field background
\begin{equation}
\label{action}
S = \frac{1}{2\pi \alpha^\prime} \int d\tau d\sigma \left( \sqrt{-\rm{det}\, g} 
+ B_{\mu \nu} \partial_\tau X^\mu \partial_\sigma X^\nu \right)\,,
\end{equation}
for an open string worldsheet with the mentioned boundary conditions. Explicitly, we want the 
boundary to define a rectangular loop in the $(X^0,X^3)$-plane with lengths
$(T,L)$. Indeed, if we want to  account for the influence of the $B$-field
we need to take a non-static configuration in which the quarks acquire
a velocity $v$ in the NC plane. We therefore take the following configuration
\begin{equation} \label{confi}
X_0 = \tau,\quad X_2=v \tau\,, \quad X_3=\sigma,\quad \rho=\rho(\sigma), 
\espai\espai -L/2<\sigma<L/2, \,\,\, 0<\tau <T \,.
\end{equation}
Plugging \bref{confi} and the NC background \bref{metrica} in the action,
we obtain
\begin{equation}
S = \frac{T}{2\pi } \int_{-{L\over 2}}^{L\over 2}
 d\sigma \left( H^{-1/2}(\rho) \left(1-\frac{v^2}{h(\rho)}\right)^{1/2}
\left( N{\rho^\prime}^2+\frac{1}{h(\rho)} \right)^{1/2} -\frac{\T}{H(\rho)-\T^2} v \right)\,,
\label{thelagr}
\end{equation}
where $\rho^\prime := \partial_\sigma \rho$ should be understood
hereafter and $H(\rho):= e^{-2 \Phi}$. In the large $T$ limit, 
the unrenormalised potential for the $q\bar{q}$ system appears as $S=T \ V_{\rm unren}$.
Notice that in the above expression there are two controllable parameters:
the noncommutativity strength $\T$ and the velocity $v$ of the quarks. 
From now on, we shall restrict to the non-supraluminical requirement $\vert v\vert < 1$,
which ensures 
$\left( 1-\frac{v^2}{h(\rho)}\right) > 0$.

\begin{figure}[t]
\begin{center}
\includegraphics[width=8cm,height=8cm]{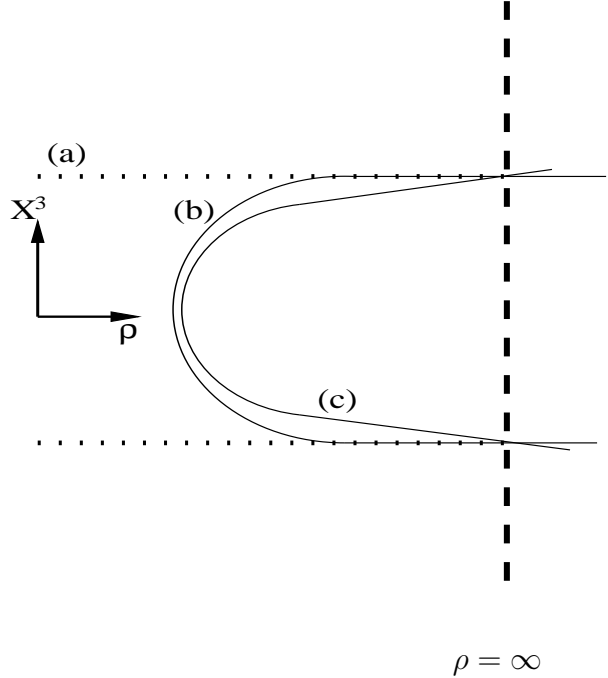}
\end{center}
\hspace{10cm} $\rho = \infty$
\caption{Different configurations for the open string worldsheet 
in the evaluation of the Wilson loop.
(a) corresponds to the
subtraction of ``two bare quarks''. 
(b) is the only
allowed configuration (fine tuned) that leads finite
 results, for both the potential and the quarks distance. (c)
is an example of a configuration that would not cancel
 the divergences at $\rho\rightarrow\infty$. The
difference between configurations (b) and (c) is 
that the first one hits the brane at right angles and,
therefore, asymptotes to (a).
}
\label{fine}
\end{figure}

We can think of the integrand for $V_{\rm unren}$ as a Lagrangian density in 
classical mechanics with $\sigma$ as the 
evolution parameter. 
Since this Lagrangian density does not depend explicitly on $\sigma$, 
its associated Hamiltonian is a conserved quantity on the extremal of
the action: 
\begin{equation}
\label{hamiltonian}
-\frac{1}{h(\rho) H^{-1/2}(\rho)} \left(1-\frac{v^2}{h(\rho)}\right)^{1/2}
\left( N{\rho^\prime}^2+\frac{1}{h(\rho)} \right)^{-1/2} + \frac{\T}{H(\rho)-\T^2} v \equiv {\rm constant}.
\end{equation}
To proceed we evaluate the constant at a special point $\rho_0$ defined as follows.
Locate the boundary of the worldsheet at some distance $\rho_{\rm max}$ from the origin, to
be sent to infinity at the end of the calculations. As we increase $\sigma$, the worldsheet
approaches the origin though the embedding $\rho(\sigma)$ until it reaches a minimum value
$\rho_0$. By symmetry of the background, this must happen at $\sigma=0$, so that
$\rho_0=\rho(0)$ and $\rho'(0)=0$. Evaluating \bref{hamiltonian} at $\rho_0$ and
solving for $\rho'$ we obtain
\begin{equation}
\label{distance}
\rho^\prime = \pm \left(\frac{H(\rho)\left[H(\rho_0)-H(\rho)\right]}{N}\right)^{1/2} \frac{1}{\alpha^2+\alpha v 
\left[\alpha^2+H(\rho_0)\right]^{1/2}+H(\rho)}\,,
\end{equation}
where we have defined the \emph{effective} or boosted noncommutative parameter as
\[
\alpha^2:=\frac{\T^2}{1-v^2}\,.
\]
Equation \bref{distance} can be used to obtain an implicit relation between
the quark separation and $\rho_0$,
\begin{equation}
\label{L}
L(\rho_0)=2\sqrt{N} \int_{\rho_0}^{\rho_{\rm max}} d\rho\, 
\frac{\alpha^2+\alpha v \left[\alpha^2+H(\rho_0)\right]^{1/2}+H(\rho)}{\left(H(\rho)\left[H(\rho_0)
-H(\rho)\right]\right)^{1/2}}\,.
\end{equation}
Similarly, we can plug equations \bref{hamiltonian} and \bref{distance} into \bref{thelagr}
to obtain a relation between the unrenormalised potential and $\rho_0$,
\be
\label{potential}
V_{\rm unren}(\rho_0) = {\sqrt{N} \over  \pi} \int_{\rho_0}^{\rho_{\rm max}}\, d\rho
\left(  \frac{\T^2 + (1-v^2) H(\rho_0)}{H(\rho)\left[H(\rho_0)-H(\rho)\right]} \right)^{1/2}.
\ee
Now, from fig. \bref{figdila}, we see that $H(\rho)$ decreases
very fast, so that $H(\rho_0) \gg H(\rho)$ for sufficiently large $\rho$.
As a consequence, (\ref{potential}) diverges as we let $\rho_{\rm max}\rightarrow \infty$,
which is interpreted as due to the presence of the two bare quark masses at
the endpoints of the string. To extract just the potential, we proceed to subtract
this contribution as usual \cite{malda-wilson,gross-wil}.
We therefore repeat the calculation for the following configuration
\begin{equation}
X_0=\tau\,,\quad X_2=v\tau\,,\quad X_3\equiv{\rm constant}\,,\quad \rho=\sigma\,,
\end{equation}
which corresponds to a straight worldsheet of a
string stretching from the initial stack of $N$ D-branes to the single one located at
infinity (see fig. (\ref{fine}.a)). Subtracting this contribution 
we obtain the following regularised quark-antiquark potential
\begin{equation}
\label{Ereg}
V_{{\rm ren}}= {\sqrt{N} \over  \pi} \left\{ \int_{\rho_0}^{\rho_{\rm max}}\, d\rho\,
\left(  \frac{\T^2 + (1-v^2) H(\rho_0)}{H(\rho)\left[H(\rho_0)-H(\rho)\right]} \right)^{1/2}
-\int_0^{\rho_{\rm
    max}} d\rho\, 
\left(\frac{(1-v^2) H(\rho)+\T^2}{H(\rho)\left[H(\rho)
+\T^2\right]}\right)^{1/2}\right\}.
\end{equation}
It is easy to check that in the commutative limit $\T=0$, both $V_{{\rm ren}}$
and $L$ remain finite as we let $\rho_{\rm max}$ grow to infinity. 
Nevertheless, arbitrary values of $\T$ require a further
restriction for the potential to be well-defined. We discuss this issue and its
physical interpretation in the next subsection.

\subsection{The fine tuning}

Consider, for a generic value of $\T$, the distance between the endpoints of the string in
the $X^3$ axis (\ref{L}). We want to keep $L$ finite as we move the boundary to  
${\rho_{\rm max}} \rightarrow \infty$. Since in this limit 
$H(\rho)\rightarrow 0$, we need
\be \label{v-con}
 \alpha^2+\alpha v \left[\alpha^2+H(\rho_0)\right]^{1/2} = 0\,.
\ee
The equation admits two solutions. The first one is $\alpha=0$, which corresponds
to the commutative case, and imposes no restrictions on $v$. This was to be expected, 
since in the absence of $B$-field, Lorentz symmetry is restored in the whole flat part
of the brane, and two quarks moving at the same velocity are equivalent to
two static quarks. Nevertheless, in the presence of a $B_{23}$, the Lorentz symmetry
is broken, and equation \bref{v-con} selects
\begin{equation}
v= - \frac{\T}{\sqrt{H(\rho_0)}}\,.
\label{finet}
\end{equation}
Since, by equation \bref{L}, $L$ determines $\rho_0$, we see that
the velocity must be fine tuned with respect to the strength of 
the $B$-field and the distance between quarks.
Remarkably, the same fine tuning reappears again when 
we consider the 
renormalised potential (\ref{Ereg}). To obtain a finite potential after the 
subtraction we need both integrands in  (\ref{Ereg}) to cancel each other when 
${\rho_{\rm max}} \rightarrow \infty$. This imposes the condition 
\be
\frac{\T^2 + (1-v^2) H(\rho_0)}{H(\rho_0)} = 1\, \espai\espai \Rightarrow \espai
\espai v^2=  \frac{\T^2}{H(\rho_0)}\,\,,
\label{finet2}
\ee
which is consistent with (\ref{finet}). Therefore, the fine tuning solves 
simultaneously the problem of fixing the distance between quarks at
the boundary at infinity,  and the problem of finiteness the potential. 
Despite being an ad hoc requirement, the fine tuning is necessary to provide
a dual supergravity interpretation of the Wilson loop in the field theory. 

The physical interpretation is somewhat analogous to the situation when a
charged particle enters a region with a constant magnetic field. In that case, 
there is also a fixed relation -say, a fine tuning- between the three relevant 
parameters: the radius of the circular orbit, the velocity, and the strength of
the magnetic field. As in our case, such a particle would not feel the
presence of the magnetic field if it did not have a non-zero velocity
transverse to it, which explains why we chose a non-static configuration
\cite{Alishahiha:1999ci}.

Notice that implementing the fine tuning in \bref{distance} shows
that now the endpoints of the string hit the boundary at ${\rho_{\rm max}} \rightarrow \infty$
at right angles, as depicted in fig. (\ref{fine}.b). This is the only way of 
keeping finite the quarks distance. For instance, the configuration
(c) in fig. \bref{fine} would not lead to a finite result.
In turn, this explains why the fine tuned configuration
allows for a finite renormalised potential, since it is the only
one that provides an asymptotic coincidence with the configuration that one 
needs to subtract.

We conclude this subsection by studying the consequences of the requirement that $v<1$.
The fine tuning demands then that
$\T^2 < H(\rho_0)$. Since $H(\rho)$ is monotonically
decreasing and tends to zero at infinity, this inequality implies two things. The first one
is that $H(0)=e^{-2\Phi_0}$ must also satisfy the inequality, so that we need $\T^2<e^{-2\Phi_0}$.
This enters in contradiction with the requirements \bref{final-condi} of section 2.3.
Therefore, to properly study the Wilson loop, we have to abandon one of the following
requirements: smallness of the dilaton, smallness of the curvature, or KK modes decoupling.
If, as in \cite{mn}, we only disregard the KK condition, we then need to impose
\be
1 \, \ll \, \T \, < \, e^{-\Phi_0} \, \ll N \,.
\ee
The second one is that $\rho_0$ has an upper bound $\rho_w$, for which $H(\rho_w)=\T^2$. 
Choosing $\rho_0>\rho_w$ would lead to supraluminical velocities
\footnote{
Having \cite{dhar-kita} in mind, we just mention
that $\rho_w$ has the property that the warp factor $e^{\Phi}h^{-1}$ in front
of the NC directions of the metric \bref{nc-mn} acquires its maximum value.}.
It is easy to see that an upper limit on $\rho_0$ implies a lower limit
on the quark separation $L$. Seeking for an understanding of this lower 
limit for $L$, it is tempting to think that this could
be related to the fact that gauge invariant objects in NC theories involve
open Wilson lines (see section 4), which exhibit a relation between
their lengths and their momentum through 
\begin{equation}
\label{incerrel}
\Delta l^{\mu}=\T^{\mu\nu}k_{\nu}\,.
\end{equation}
In our case the length $L$ is along $X^3$ whereas the velocity is along $X^2$,
in agreement with our NC parameter $\T^{23}$. A complementary consideration \cite{susskind}
is that relation (\ref{incerrel}) gives the size of the particle in the $X^3$ direction when
it has a given momentum along $X^2$. However all these are still vague
arguments that we do not claim as conclusive.

\subsection{The results}
\begin{figure}[t]
\begin{center}
\includegraphics[width=7cm,height=7cm]{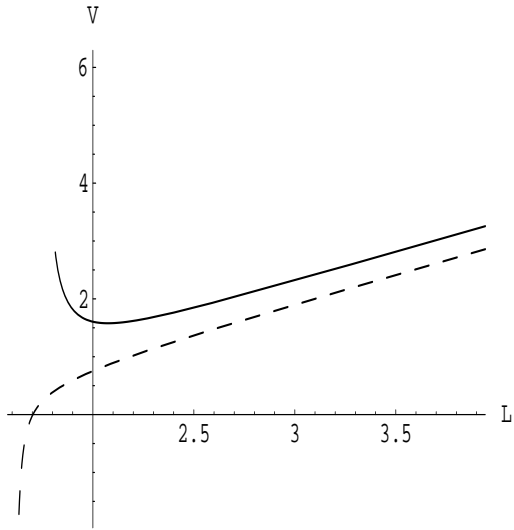}
\end{center}
\hspace{0.5cm}
\caption{Quark-antiquark
potential versus their separation. The dashed line
corresponds to the commutative case, while the full curve depicts the corresponding NC one.
At large distance, both theories confine, while as we move the quarks closer, the UV 
physics give a completely different behaviour.}
\label{fig2}
\end{figure}

Once the necessity for the fine tuning has been discussed, we proceed
to apply it to our formulas \bref{L} and \bref{Ereg} to obtain the
simplified expressions for the quarks distance and the renormalised
potential:
\begin{equation}
\label{L2}
L=2\sqrt{N} \int_{\rho_0}^{\infty} d\rho 
\left(\frac{H(\rho)}{H(\rho_0)-H(\rho)}\right)^{1/2}\,,
\end{equation}

\begin{equation}
\label{pot}
V_{\rm ren} = \frac{\sqrt{N}}{\pi}  
\left\{ \int_{\rho_0}^{\infty} d\rho\,
\sqrt{\frac{H(\rho_0)}{H(\rho)\left[H(\rho_0)-H(\rho)\right]}}-\int_0^{\infty} d\rho\, 
\sqrt{\frac{{\T^2\over H(\rho_0)}\left[H(\rho_0)-H(\rho)\right] +  H(\rho)}
{H(\rho)\left[\T^2 + H(\rho)\right]}}
\right\}\,.
\end{equation}
Both equations can be used to obtain $V_{\rm ren}$ as a function of $L$. Although the
relation cannot be given algebraically, one can make a numerical plot to
study the phases of the theory. In fig.\bref{fig2}, we present
the plot of the renormalised potential against the distance between quarks
in both the commutative (dashed line) and the NC (full line) backgrounds.

The first immediate observation is that both theories exhibit the same behaviour
in the IR. At large separation, the potential is linear in both cases and, restoring
$\alpha'$ factors, we obtain
\be
V_{\rm ren}(L) \, \approx \, {e^{\Phi_0}\over 2\pi \alpha'}\,L \,,
\ee
independent of the value of $\T$. Indeed, this result 
can be proven analytically, and does not rely only on the numerical analysis.

Nevertheless, as we move the quarks closer, the two theories exhibit
a very different behaviour. In the NC case, the potential becomes extremely
repulsive, presumably due to the expected effects of the noncommutative
uncertainty relations at short distances. On the other hand, the commutative
potential starts deviating from the linear behaviour in the opposite way,
although this happens in a region where the commutative dilaton \bref{com-dilato}
is not small anymore, and so the calculation should have been continued in the NS5 S-dual
picture.

\section{Gauge theory physics from noncommutative MN}
\setcounter{equation}{0}

In this section we try to extract more information of the noncommutative
gauge theory from the proposed supergravity dual. Our discussion will
be parallel to that in \cite{lerda,zaffaroni}, where they studied
the commutative Maldacena-\nun solution. 
We will follow the conventions of \cite{lerda}

\subsection{NC Yang-Mills coupling as a function of $\rho$}


Let us then begin with the discussion on the Yang-Mills coupling for the commutative
case. 
The proposal in
\cite{lerda} is that one can obtain $g_{\rm YM}$ as a function of $\rho$ by the following procedure.
Consider the DBI action of a D5 in the background of MN. Take the $\a'\rightarrow 0$
limit and promote the abelian fields to transform in the adjoint of $SU(N)$.
That would give a $SU(N)$ Yang-Mills action in the curved space that the
D5 are wrapping, which in our case is $\MR^4 \times S^2$. 
Since we are interested in the IR of the gauge theory, we take a limit in which
the volume of the $S^2$ is small, so that the action becomes, upon an $S^2$ reduction,
a four dimensional \CN=1 $SU(N)$ SYM with the following bosonic structure
\be \label{Action}
S[A_{\mu}]=-{1\over 4g_{\rm YM}^2}\int_{\MR^4} d^4x \, \, F_{\a\b}^A F_{A}^{\a\b}\,, 
\espai \espai \espai \a,\b=0,1,2,3\,.
\ee
Indeed, one would get a series of corrections from the KK modes of the $S^2$
which, as discussed in section \ref{sugra}, 
decouple under a certain choice of $N$, $\T$ and $\Phi_0$. 
The YM coupling
appearing in \bref{Action} is essentially given by the inverse volume of the
$S^2$ measured with the ten-dimensional commutative metric, and it depends
on the radial coordinate $\rho$.\footnote{In the original paper of Maldacena-\nun
the YM coupling was calculated directly in the gauged supergravity.
At the end of the day, it differed from the one in \cite{lerda} by the fact
that the volume of the $S^2$ was calculated with the seven-dimensional metric.
The remarkable matching of \cite{lerda} with the field theory result seems
to select their method.}

Let us first adapt this method in order to obtain
$\hat{g}_{\rm YM}(\rho)$ for the NC-MN solution \bref{nc-mn}. We should 
now expand the DBI including the background $B$-field. Actually, in the low-energy
limit, it was shown in \cite{sei-wit} that the theory becomes noncommutative.
When the dilaton is independent of the gauge theory coordinates, \ie at zero momentum,
the DBI action with a constant magnetic $B$-field gives,
in the low-energy limit, 
the same quadratic terms as its noncommutative version
\be \label{nc-DBI}
S_{\rm DBI}[\hA_{\mu}]={\tau_5 \over  G_o} \int_{\MR^4\times S^2} d^6x
\,\,\sqrt{\det\left(f^*G+2\pi\hF
\right)_{\star}} 
\ee
where $G_o$, $\hF$ and $G_{\mu\nu}$ are the effective coupling constant, field 
strength and metric seen by the open strings in a $B$-field background. Furthermore, $\tau_5$ 
denotes the string tension.
The pullbacks to the brane world-volume are denoted by $f^*$.
All products in \bref{nc-DBI} are understood as Moyal $\star$-products with noncommutative
parameter $\T^{\mu\nu}$.
The relations between the open string quantities and the
closed string ones $e^{\hat{\Phi}}$, $F$ and $g_{\mu\nu}$ are
\begin{eqnarray} \label{sw-map}
G_{\mu\nu}=g_{\mu\nu}- (f^{*}B)_{\mu\rho} g^{\rho\lambda}(f^*B)_{\lambda\nu}\,,
&&\espai\espai \hF={1\over 1+ F\T}\,F\,, \nonumber\\
G_o=e^{\Phi}\left({ \det G \over \det g}\right)^{1/4}\,, \espai\espai\espai &&
\T^{\mu\nu}=-g^{\mu\rho} (f^{*}B)_{\rho\lambda} G^{\lambda\nu}\,.
\end{eqnarray}
In order to correctly identify the $\hg_{\rm YM}$ for the noncommutative theory,
we use the noncommutative action and variables. Expanding \bref{nc-DBI}
and plugging in our background \bref{nc-mn} we obtain
\be \label{nc-action}
S[\hA_{\mu}]=-{1\over 4\hg_{\rm YM}^2}\int_{\MR^4} d^4x \, \, \hF_{\a\b}^A \star
\hF_{A}^{\a\b}\,, 
\espai \espai \espai \a,\b=0,1,2,3
\ee
with the following expression for the noncommutative YM coupling
\be
{1\over \hg^2_{\rm YM}(\rho)}= {2\pi^2 \tau_5 \over N^2 G_o} e^{2\Phi(\rho)}
e^{-4g(\rho)}  \int_{S^2} d\t d\phi \sqrt{f^*G}\,.
\ee
By explicit calculation, it turns out 
that the
Yang-Mills coupling can be written in the following way
\be
{1\over \hg^2_{\rm YM}(\rho)}={N\over 32 \pi^2} Y(\rho) \int_0^\pi \, d\t \,\,
 \sin\t \left[ 1+{\cot^2 \t\over Y(\rho)}\right] ^{\undos}
\ee
where we defined
\be
Y(\rho)=4\rho \coth 2\rho -1\,.
\ee
By comparison with \cite{lerda}, we see that the relation between $\hg_{\rm YM}$ and 
the radial coordinate turns out to be identical to that of $g_{\rm YM}$!

\subsection{Relation between $\rho$ and the energy}

To go further and obtain the $\b$-function, we still need to find the
relation between $\rho$ and the energy scale of the dual field theory.
In the commutative MN, there are two basic lines of argument that lead
to the same conclusions. We briefly review them in order to be applied
to the noncommutative case.

\avall

({\it i}) The authors of \cite{lerda} observe that \CN=1 SYM theories
have a classical $U(1)_R$ symmetry which is broken at the quantum
level (and after considering non-perturbative effects) to $Z_2$. 
An order parameter is the vacuum expectation value of
the gaugino condensate $<\l^2>$, i.e. if $<\l^2>\ne 0$,
the symmetry is broken. To relate this phenomenon to the supergravity
side, one is guided by the fact that we know how the $U(1)_R$ symmetry
acts, since it simply corresponds to rotations along the angle $\psi$.
It is easy to realise that such rotations are an isometry of the metric
if and only if the supergravity field $a(\rho)$ appearing in \bref{nc-mn}
is zero. Therefore one is led to conjecture that $a(\rho)$ is the
supergravity field dual to the gaugino condensate. The argument finishes by
noticing that since $<\l^2>$ has protected dimension three, it must
happen that \footnote{The proportionality coefficient is 1 from 
explicit calculations \cite{hollowood}.}
\be
<\l^2>=\Lambda^3
\ee
where $\Lambda$ is the dynamically generated scale.
This leads to the following implicit relation between $\rho$ and
the field theory scale $\mu$
\be
\label{afield}
a(\rho) \propto {\Lambda^3\over \mu^3}\,.
\ee

\avall

({\it ii}) A slightly different argument is given in \cite{zaffaroni}.
The authors 
first expand $g_{\rm YM}(\rho)$ for large $\rho$ (in the UV)
where it can be compared to perturbative results of the gauge theory,
\ie with $g_{\rm YM}(\mu/\Lambda)$. This immediately gives the searched relation
$\rho=\rho(\mu/\Lambda)$, valid in the UV region. Indeed, they also identify
$a(\rho)$ as dual to the gaugino condensate by 
trying to guess what is the exact form of the mass term for the
gauginos in the four-dimensional \CN=1 SYM. Gauge invariance
of the Lagrangian must involve couplings to the gauge field through covariant
derivatives. This fact, together with the detailed knowledge of how the twisting
of the field theory is performed, allowed the authors to find that the Lagrangian
must involve a term like
\be
a(\rho) \, \bar{\l} \l \,.
\ee
Applying standard arguments of the original AdS/CFT correspondence one would conclude
again that $a(\rho)$ is the supergravity field dual to the gaugino condensate.

\avall

We now try to adapt these arguments to our NC-MN solution. The first important remark is that
noncommutative gauge theories do not have local gauge-invariant operators 
\cite{gross, rey}.
Terms like $tr (\hF_{\mu\nu}\star \hF^{\mu\nu})$ are only gauge invariant after integration
over all the space. This fact increases the difficulty to associate the dual supergravity fields, 
since they should act as sources of gauge-invariant operators.
Nevertheless, since translations are still a symmetry of the
theory, there must exist gauge-invariant operators local in momentum space. Such operators
involve the so-called open Wilson lines, whose length must
be proportional to the transverse momentum. Explicitely, if we name $\Delta l^{\mu}$
the separation between the endpoints of an open Wilson line, and $k_{\mu}$ its momentum
in the noncommutative directions, gauge-invariance requires
\be \label{nc-relation}
\Delta l^{\mu}=\T^{\mu\nu}k_{\nu}\,.
\ee
Several scattering computations \cite{liu1,okawa,das} seem to 
confirm that a general supergravity field $h$ 
couples to the noncommutative version of the ordinary operator to which it coupled
when $\T=0$  via
\be
\int d^dk \, h(-k) \hat{\cO}(k)\,.
\ee
The noncommutative operator $\hat{\cO}(k)$ is defined from its commutative local one $\cO(x)$ 
by inserting the mentioned Wilson line $W[x,\cC]$ and Fourier-transforming\footnote{
We refer to {\it e.g.} \cite{gross, rey, liu1} for further discussions on the
ambiguity of the insertion of $\cO(y)$ along the contour $\cC$, and for general aspects
of open Wilson lines. We shall only make use of a few of their properties.}
\be
\hat{\cO}(k)=tr P_{\star} \int d^dx \, [W(x,\cC) \cO(y)]\star e^{ikx}
\ee
where $\cC$ is a straight path connecting the endpoints separated according to \bref{nc-relation}
and $y$ is an arbitrary point along $\cC$.

The observation is that the relevant fields appearing in our background \bref{nc-mn} do not depend
on the noncommutative coordinates $(x^2,x^3)$, so that their Fourier-transforms would
involve a delta function in momentum space. In other words, we just need the zero-momentum
couplings, where the length of the Wilson lines vanishes, and $\hat{\cO}$ reduces to $\cO$.

We are now ready to apply the arguments ($i$) and ($ii$) to our case. As far as $U(1)_R$
symmetry breaking in the supergravity solution is concerned, nothing changes 
with respect to the commutative case. Again, shifts of $\psi$ are an isometry of the NC metric
if and only if $a(\rho)=0$. This is due to the fact that the only change in the metric
is a factor of $h^{-1}(\rho)$ in front of $\, dx_{2,3}^2$.

For the same reason, 
the whole structure of the twisting of the normal bundle to the $S^2$ inside
the Calabi-Yau threefold is also unchanged. At zero-momentum in the noncommutative directions,
gauge invariance in the field theory demands again that the fermionic couplings to
the gauge fields appear only via covariant derivatives. So it looks like the arguments
of ($i$) and ($ii$) lead again to conjecture that $a(\rho)$ is dual to the gaugino condensate.

Indeed, independently of this relation, one could proceed as in \cite{zaffaroni} and
expand $\hg_{\rm YM}(\rho)$ at very large $\rho$. In that region, the theory is in the UV 
and perturbative calculations should be trustable. The field theory results 
(see the review \cite{szabo} and references therein) show 
that the perturbative NC $U(N)$ $\b$-function is identical to the $SU(N)$ commutative 
one.\footnote{Recall that the $U(1)$ degrees of freedom inside a noncommutative $U(N)$
gauge theory do not decouple and, unlike the commutative case, they run with
the same $\b$-function as the rest of the $SU(N)$ \cite{Armoni}.} So both the supergravity 
behaviour of $\hg_{\rm YM}(\rho)$ and the perturbative behaviour of the NC $\b$-function
are identical to the commutative case. The conclusion is that the relation
between $\rho$ and $\Lambda/\mu$ is also unchanged.

Summarising, it seems like the NC $\b$-function calculated from \bref{nc-mn}
and the commutative one extracted from the commutative MN are identical.
Hence
the same results found in \cite{lerda,zaffaroni} hold in our case. We just recall that
properly choosing the proportionality function in (\ref{afield})
\cite{olesen, sannino}  
one remarkably obtains the whole perturbative NSVZ $\b$-function 
\be
\beta(g_{\rm YM}) = -\frac{3g^3_{\rm YM} N}
{16\pi^2}\left[1-\frac{N g^2_{\rm YM}}{8 \pi^2}\right]^{-1}\,.
\ee
and they even identified 
the contribution of (presumably) non-perturbative 
fractional instantons.


\appendix
\setcounter{equation}{0}
\newcounter{zahler}
\addtocounter{zahler}{1}
\renewcommand{\thesection}{\Alph{zahler}}
\renewcommand{\theequation}{\Alph{zahler}.\arabic{equation}}
\label{appdx}

\setcounter{section}{0}
\setcounter{subsection}{0}

\renewcommand{\thesection}{\Alph{zahler}}
\renewcommand{\theequation}{\Alph{zahler}.\arabic{equation}}

\setcounter{equation}{0}
\setcounter{zahler}{0}
\addtocounter{zahler}{1}
\renewcommand{\thesection}{\Alph{zahler}}
\renewcommand{\theequation}{\Alph{zahler}.\arabic{equation}}

\section{Appendix}
In this appendix we collect the conventions and definitions
necessary to read the Maldacena-\nun background in \bref{metrica}-\bref{com-dilato},
referring the reader to the original references for a careful derivation.

There are two functions of the radial variable $\rho$, which are given by
\be 
e^{2g(\rho)}=\rho \coth 2\rho - {\rho^2 \over \sinh^2 2\rho} -{1\over 4}\,,
\espai\espai\espai
\label{A}
a(\rho)={2\rho \over \sinh 2\rho}\,.
\ee
The $SU(2)$ gauge-field $A$ is parametrised by
\be
A=\undos\left[\s^1 a(\rho) d\theta+\s^2 a(\rho) \sin\t d\phi + \s^3 \cos\t d\phi\right]\,,
\ee
where
\be
A=A^a{\s^a\over 2}\,, \espai\espai F=F^a {\s^a\over 2}\,, \espai\espai 
F^a_{\mu\nu}=\partial_{\mu}A^a_{\nu}-\partial_{\nu}A^a_{\mu}
+\epsilon^{abc}A^b_{\mu}A^c_{\nu}
\ee
should be understood.
Finally, the $SU(2)$ left-invariant one forms parametrising the transverse $S^3$ are
\be
w^1+i w^2 = e^{-i\psi}\left( d\t_1 +i \sin d\phi_1 \right), \espai\espai\espai
w^3=d\psi+cos\theta_1 d\phi_1.
\ee

\vskip 6mm

{\it{\bf Acknowledgements}}

We are indebted to Joaquim Gomis for the initial guidance, useful discussions
and permanent support. We are also grateful to Alberto Lerda, David Mateos, 
Carlos N\'u\~nez, Stathis Pakis, Joan Sim\'on and Paul Townsend for useful discussions.
T.M. is supported by a grant from the Commissionat per
a la Recerca de la Generalitat de Catalunya.

\end{document}